Which of the world's institutions employ the most highly cited researchers? An analysis of the data from highlycited.com

Lutz Bornmann* & Johann Bauer**

* corresponding author:

Division for Science and Innovation Studies

Administrative Headquarters of the Max Planck Society

Hofgartenstr. 8,

80539 Munich, Germany.

Email: bornmann@gv.mpg.de

** Max Planck Institute of Biochemistry

Am Klopferspitz 18,

82152 Martinsried, Germany.

EMail: jbauer@biochem.mpg.de

**Abstract**

A few weeks ago, Thomson Reuters published a list of the highly cited researchers worldwide (highlycited.com). Since the data is freely available for downloading and includes the names of the researchers' institutions, we produced a ranking of the institutions on the basis of the number of highly cited researchers per institution. This ranking is intended to be a helpful amendment of other available institutional rankings.

In 2014, Thomson Reuters published a list of the highly cited researchers worldwide (highlycited.com). This list was also published in a report (Thomson Reuters, 2014) and the data can be downloaded as an Excel file for further statistical analyses. Thus Myklebust (2014), for example, used the data to undertake a breakdown by country of the distribution of the researchers.

In order to identify the highly cited researchers, Thomson Reuters (provider of the Web of Science) selected, in a first analysis step, the publications from the natural and social sciences with document type "article" and "review" and publication years between 2002 and 2012. Then, they determined those publications which belonged to the top 1% by citations in their subject area and publication year. In a second analysis step, the authors of these highly cited publications were sorted by the discipline (e.g. Materials Science; see http://in-cites.com/thresholds-citation.html).[1] In a third analysis step, a ranking was set up within a discipline: The more highly cited publications there were for a researcher, the higher his or her rank in the discipline. In the list of the highly cited researchers published in the URL above and the report mentioned above, those researchers are listed whose rank is less than or equal to the square root of the population consisting of all researchers in a discipline with at least one highly cited publication. A total of 3215 researchers appear in the list of highly cited researchers. These are 3215 rows representing appearances of researchers and their institutions because of selection in one or more disciplines. Apparently the actual number of unique researchers is 3073.

In this study, we investigated the global distribution of highly cited researchers across institutions. For this evaluation, an elaborate cleaning process was necessary, since many institutions were not consistently named by their authors, but with several variants of their name. In this cleaning process we also combined all individual institutions of an organization – insofar as they could be recognized. Thus, for example, we combined all individual universities of the University of California system and all Max Planck institutes of the Max Planck Society. Unfortunately, some of the institutions named by the researchers could not be processed by us, since the data was too unspecific (e.g. USA) or ambiguous (an abbreviation can stand for more than one institution, even within a country).

---

[1] Disciplines are sets of journals in which the highly cited publications appeared.

Many highly cited researchers mentioned not just one, but up to five different institutions. For this reason, we produced three ranking lists, which include these institutions in different ways. The first ranking list of institutions (see Table 1) is based on the first-named institution for each researcher (his or her primary institution). Corresponding to <u>primary institutions</u>, the largest number of highly cited researchers work at the University of California (n=179). This is followed by Harvard University (n=107).

Table 1. Numbers of highly cited researchers per institution, determined by their primary institution. The 20 institutions with the highest number of highly cited researchers are shown.

| Rank | Primary institution of a researcher | Number |
|---|---|---|
| 1 | University of California, USA | 179 |
| 2 | Harvard University, USA | 107 |
| 3 | National Institutes of Health (NIH), USA | 91 |
| 4 | Stanford University, USA | 56 |
| 5 | Max Planck Society, Germany | 52 |
| 6 | Chinese Academy of Sciences, China | 46 |
| 7 | University of Texas, USA | 43 |
| 8 | University of Oxford, UK | 33 |
| 9 | Duke University, USA | 32 |
| 9 | Massachusetts Institute of Technology (MIT), USA | 32 |
| 11 | University of Michigan, USA | 31 |
| 12 | University of London, UK | 30 |
| 12 | Wellcome Trust Sanger Institute, UK | 30 |
| 14 | Broad Institute, USA | 28 |
| 14 | EMBL, UK - Germany | 28 |
| 14 | Northwestern University, USA | 28 |
| 17 | Princeton University, USA | 27 |
| 17 | University of Washington, USA | 27 |

| 19 | Brigham & Women's Hospital, USA | 26 |
| 19 | Johns Hopkins University, USA | 26 |

The second ranking list of institutions (see Table 2) is based on all the institutions named by a highly cited researcher. The evaluation with all the named institutions leads to an interesting change in the ranking list. Compared with Table 1, the ranking order of the institutions hardly changes in the higher positions; but now King Abdulaziz University appears in second place. Apparently, a great number of researchers mention this institution as an additional institution besides their primary one.

Table 2. Numbers of highly cited researchers per institution taking into account all the institutions mentioned by a researcher. The 20 institutions with the highest numbers of highly cited researchers are shown.

| Rank | All of a researcher's named institutions | Number |
|---|---|---|
| 1 | University of California, USA | 198 |
| 2 | King Abdulaziz University, Saudi Arabia | 160 |
| 3 | Harvard University, USA | 146 |
| 4 | National Institutes of Health (NIH), USA | 97 |
| 5 | Stanford University, USA | 60 |
| 6 | Max Planck Society, Germany | 57 |
| 7 | Chinese Academy of Sciences, China | 48 |
| 8 | Massachusetts Institute of Technology (MIT), USA | 44 |
| 8 | University of Texas, USA | 44 |
| 10 | University of Oxford, UK | 37 |
| 11 | University of London, UK | 35 |
| 11 | Wellcome Trust Sanger Institute, UK | 35 |
| 13 | Broad Institute, USA | 34 |
| 14 | Duke University, USA | 32 |
| 14 | University of Michigan, USA | 32 |
| 16 | EMBL, UK - Germany | 31 |
| 16 | University of Washington, USA | 31 |
| 18 | Johns Hopkins University, USA | 30 |
| 18 | Northwestern University, USA | 30 |
| 20 | Princeton University, USA | 29 |

As Bhattacharjee (2011) reported some years ago in *Science*, Saudi Arabian universities offer highly cited researchers contracts in which the researchers commit themselves to listing the Saudi Arabian university as a further institution in publications (or on highlycited.com). In return, the researchers receive an adjunct professorship which is connected with an attractive salary and a presence at the University of only one or two weeks per year (for teaching duties on site). Gingras (2014a) names the added institutions as "dummy affiliations, with no real impact on teaching and research in universities, allow marginal institutions to boost their position in the rankings of universities without having to develop any real scientific activities."

Table 3. Numbers of highly cited researchers per institution using the fractionated method. The 20 institutions with the highest numbers of highly cited researchers are shown.

| Rank | All of a researcher's named institutions | Number |
| --- | --- | --- |
| 1 | University of California, USA | 178.00 |
| 2 | Harvard University, USA | 110.50 |
| 3 | National Institutes of Health (NIH), USA | 93.00 |
| 4 | King Abdulaziz University, Saudi Arabia | 80.28 |
| 5 | Stanford University, USA | 55.50 |
| 6 | Max Planck Society, Germany | 49.50 |
| 7 | Chinese Academy of Sciences, China | 41.33 |
| 8 | University of Texas, USA | 39.50 |
| 9 | Massachusetts Institute of Technology (MIT), USA | 33.08 |
| 10 | University of Oxford, UK | 32.08 |
| 11 | Wellcome Trust Sanger Institute, UK | 31.33 |
| 12 | University of Michigan, USA | 30.83 |
| 13 | Duke University, USA | 29.50 |
| 14 | University of London, UK | 29.33 |
| 15 | University of Washington, USA | 29.00 |
| 16 | Princeton University, USA | 27.33 |
| 17 | EMBL, UK - Germany | 27.17 |
| 18 | Northwestern University, USA | 26.50 |
| 19 | Johns Hopkins University, USA | 26.25 |
| 20 | University of Cambridge, UK | 23.67 |

Many researchers listed only one institution, but others two or more (up to five). These institutions can either be counted as units (as in Table 2) or as fractions. In the fractionated method, the number of institutions listed by a researcher is taken into account: If he or she has listed three institutions, for instance, each institution is counted as 1/3. The result of the fractionated method for the number of highly cited researchers per institution is given in Table 3. As expected, the number of highly cited researchers at King Abdulaziz University is especially reduced (from 160 to about 80).

Comparing tables 1, 2, and 3 it is interesting to see that the first six positions – with the obvious exception of King Abdulaziz University – are the same in the three rankings, while the lower ones are scrambled. That is because the differences between the institutions are generally larger at the beginning and smaller at the end. It is clearly visible in all three tables that the differences in the number of highly cited researchers are relatively large for institutions at the first six positions. This leads to robust findings although different counting methods are used in the three tables. Since the differences of the number of highly cited researchers at the lower positions are relatively small, the different counting methods lead to different institutional positions.

The evaluation of the list of highly cited researchers on the basis of institutions is an interesting alternative to the usual university rankings (such as, for example, the Leiden Ranking, leidenranking.com). Since scientific work is performed by individuals and the attribution of success is generally applied on the level of the individual (such as via the Nobel Prize) (Ziman, 2000), counting the number of successful persons seems more reasonable than counting the number of successful publications (as with the Leiden Ranking's number of highly cited publications per institution). To be most cited as a scientist means to be well known or to be attractive for discussions, and that made the citation indexes in the last centuries so attractive and important, for scientists and bibliometricians. However, it does not mean automatically that highly cited authors have produced high quality research. For example, it was found by Garfield (2006) that Nobel Prize winners are often highly cited, but by far not all highly cited authors are Nobel Prize winners.

The results for King Abdulaziz University illustrate that university rankings can be manipulated. In a similar analysis of the highly cited researchers dataset Gingras (2014b) concluded: "All these data certainly suggest that this particular institution has found a cheap way to be considered 'excellent' in world university rankings". A manipulation of the list of highly cited researchers has also consequences for the Academic Ranking of World Universities (ARWU, http://www.shanghairanking.com/) – the oldest and best-known international university ranking (Hazelkorn, 2011). ARWU considers data from highlycited.com to rank universities according to their number of Highly Cited Researchers in 21 subject categories. "These individuals are the most cited within each category. If a Highly Cited Researcher has two or more affiliations, he/she was asked to estimate his/her weights (or number of weeks) for each affiliation. More than 2/3 of those multi-affiliated Highly Cited Researchers provided such estimations and their affiliations receive the weights accordingly. For those who did not answer, their first affiliation is given a weight of 84% (average weight of the first affiliations for those who replied) and the rest affiliations share the remaining 16% equally" (http://www.shanghairanking.com/ARWU-Methodology-2013.html).

To counteract attempts at manipulation, ARWU should only consider primary institutions of highly cited researchers.

# Acknowledgements

We would like to thank David Pendlebury from Thomson Reuters for recommendations to improve the manuscript. The comparison of our lists of highly cited researchers per institution with his lists led to very similar results. Initial differences between the lists have been resolved accordingly.